\documentstyle[epsfig,amsmath]{mn}
\def\spose#1{\hbox to 0pt{#1\hss}}
\def\lta{\mathrel{\spose{\lower 3pt\hbox{$\mathchar"218$}}\raise 2.0pt\hbox{$\mathchar"13C$}}}
\def\gta{\mathrel{\spose{\lower 3pt\hbox{$\mathchar"218$}}\raise 2.0pt\hbox{$\mathchar"13E$}}}

\def\om{\Omega_{0}}
\def\oml{\Omega_{\Lambda}}

\title[K-z Relations for Luminous Radio Galaxies]
{Infrared Magnitude-Redshift Relations for Luminous Radio Galaxies}
      
\author[K.\,J.\, Inskip {\rm et al.}]
{K.\,J.\, Inskip$^1$\footnotemark, P.\,N.\,Best$^{2,3}$,
M.\,S.\,Longair$^1$ and D.\,J.\,C.\,MacKay$^1$\\
$^1$ Cavendish Astrophysics Group, Cavendish Laboratory, Madingley Road, Cambridge, CB3 0HE,\\ 
$^2$ Sterrewacht Leiden, Postbus 9513, 2300 RA Leiden, the Netherlands\\
$^3$ Institute for Astronomy, Royal Observatory Edinburgh, Blackford Hill, Edinburgh, EH9 3HJ\\}

\date{ }

\pagerange{\pageref{firstpage}--\pageref{lastpage}}

\pubyear{2001}

\begin{document}

\label{firstpage}
\maketitle
\begin{abstract}
Infrared magnitude-redshift relations (K-z relations) for the 3CR and 6C
samples of radio galaxies are presented for a wide range of plausible
cosmological models, including those with non-zero cosmological
constant $\oml$. Variations in the galaxy formation redshift,
metallicity and star formation history are also considered.  The
results of the modelling are displayed in terms of magnitude
differences between the models and no-evolution tracks, illustrating
the amount of K-band evolution necessary to account for the
observational data.    

Given a number of plausible assumptions, the results of these analyses
suggest that: (i) cosmologies which predict ${T_0}\times{H_0} \gta 1$
(where $T_0$ denotes the current age of the universe) can be
excluded; (ii) the star formation redshift should lie in the redshift
interval $5\leq z_{\rm f} \leq 20$, values towards the lower end of
the range being preferred in cosmologies with larger values of
${T_0}\times{H_0}$; (iii) the Einstein-de Sitter model ($\om = 1,\,\oml = 0$)
provides a reasonable fit to the data; (iv) models with finite values
of $\oml$ can provide good agreement with the observations only if
appropriate adjustments of other parameters such as the galaxy
metallicities and star-formation histories are made. Without such
modifications, even after accounting for stellar evolution, the high
redshift radio galaxies are more luminous (that is, more massive) than
those nearby in models with finite $\oml$, including the favoured model
with $\om = 0.3,\,\oml = 0.7$. For cosmological
models with larger values of ${T_0}\times{H_0}$, the conclusions are the same
regardless of whether any adjustments are made or not. 
The implications of these results for cosmology and models of galaxy
formation are discussed.

\end{abstract}


\section{Introduction}

\footnotetext{E-mail: kji@mrao.cam.ac.uk}

The models used to account for the infrared magnitude-redshift
relation of luminous radio galaxies are sensitive to the assumptions
made concerning the evolution of their stellar content, as well as the
choice of cosmological model.  Galaxies at a particular redshift are
observed at different cosmic epochs in different cosmological models
and so their predicted spectra at a given redshift can vary quite
dramatically.   Accurate modelling of the evolution of galaxy spectra
with cosmic time is required if the infrared K magnitude-redshift
relation is to be used to constrain the range of acceptable world
models.  In this paper the spectral evolution of galaxies is modelled
for different plausible cosmological models using the most recent
spectral synthesis codes of Bruzual and Charlot (2001), \textsc{gissel}2000,
and these results compared with the observational data.  Even for the
simplest cases, numerous cosmological and evolutionary effects need to
be taken into account and it is often not intuitively obvious how
changes in the assumptions made change the predicted relations -- one
of the objectives of this paper is to provide these insights. Although
our prime concern is the K-z relation for radio galaxies, the
considerations of this paper are relevant to all such cosmological
studies.  

The radio galaxies associated with the most powerful extragalactic
radio sources seem to form a remarkably uniform class of galaxy out to
large redshifts.  Following the early work of Lilly and Longair
(1984), Best {\it et al.} (1998) carried out a much more detailed
analysis of the photometric properties of a complete sample of 3CR
radio galaxies, the brightest radio galaxies in the northern sky
selected at 178 MHz (Laing {\it et al.} 1983). They modelled the K-z
relation for world models with $\om = 0$ and $\om = 1$, in both cases
assuming $\oml = 0$.  These models included passive evolution of the
stellar populations of the galaxies with formation redshifts of 3, 5
and 20. The passive evolution models in the $\om = 1$ universe
provided a reasonable fit to the observations, whereas those with $\om
= 0$ gave a poor fit.  The models with low redshifts of formation,
$z_{\rm f}\approx 3$, resulted in more luminous galaxies than those
observed at $z \ga 1.2$, suggesting that the stellar populations of
the galaxies must have formed at $z > 3$.  

Eales and Rawlings (1996) and Eales {\it et al.} (1997) determined the
K-z relation for a 
complete sample of 6C radio galaxies which were selected at 151 MHz
and are about six times fainter in flux density than the 3CR sample
(Eales 1985).  The 6C host galaxies have similar K-luminosities to
those of the 3CR galaxies at $z \lta 0.8$, but are about 0.6 magnitudes
fainter at larger redshifts. Nevertheless, the high redshift 6C
galaxies are significantly brighter than expected in world models in
which no evolution of the stellar populations of the galaxies is
assumed.  

The objectives of this paper are to review the current status of the
K-z relation for 3CR and 6C luminous radio galaxies (Section 2) and
then to carry out a more extensive analysis of the resulting K-z
relation for a wider range of plausible world models than those
considered by Best {\it et al.} (1998) and Eales {\it et al.} (1997),
including models with a non-zero 
cosmological constant.  The predicted evolution of the stellar
populations of the galaxies is derived from the most recent spectral
synthesis codes of Bruzual \& Charlot (1993, 2001),
\textsc{gissel}2000; a wide range of stellar synthesis models is
considered, as well as the effects of changes in metallicity. The 
details of the different cosmological models, the spectral synthesis
codes and procedures used to predict the K-z relations are described
in Section 3.  The results of these calculations are described in
Section 4, indicating which parameters have the greatest impact upon
the form of the predicted K-z relation.  Also included in this Section
are the results of varying the star formation history and metallicity
in plausible ways.  Most massive galaxies at small redshifts have high
metallicities and colour-magnitude relations show that more massive
galaxies are redder in colour, suggesting a mass-metallicity
correlation (see, for example, Arimoto \& Kodama, 1996). On the other
hand, at high redshifts, we might expect a lower metal content, as
suggested by the analyses of Pei {\it et al.} (1999) and Pettini {\it
et al.} (1997).  In Section 5 the results of these analyses are
discussed and our conclusions are summarised in Section 6.
Throughout the paper, $H_0 = 50$ km s$^{-1}$ Mpc$^{-1}$ is adopted,
unless otherwise stated.   
 
\section{The K-z Relation for Luminous Radio Galaxies}

\subsection{The Revised K-z Diagram}

The observational data used in this analysis consist of infrared K
magnitudes for 72 3CR galaxies in the redshift interval $0 < z< 1.8$
(Best {\it et al.} 1998), and 57 6C galaxies in the range $0 <z <3.4$
(Eales {\it et al.} 1996, 1997). The 3CR galaxies comprise a complete
sample of radio sources with flux densities $S_{178} \ga 10.9$ Jy at 178
MHz, with a mean flux density of about 15 Jy; the 6C sample is
complete in the flux density interval $2.2 \le S_{151} \le 4.4$ Jy at
151 MHz, with an equivalent mean flux density at 178 MHz of about 2.6
Jy.  The 3CR K magnitudes were obtained for $z < 0.6$ from
observations with the 
\textsc{ircam3} array on the UK Infrared Telescope (Best {\it et al.}
1997), and for $z < 0.6$ from the photometry of Lilly \& Longair
(1984).  The 6C data points were mostly derived from \textsc{ircam}
observations, but also from colour corrected K$^\prime$ observations
with the \textsc{redeye} camera of the Canada-France-Hawaii Telescope
(Eales {\it et al.} 1997).   

Various factors had to be included to create a self-consistent set of
K magnitudes.  Firstly, aperture corrections were made, reducing the
magnitudes to a standard circular aperture of metric diameter
63.9\,kpc. These aperture corrections are dependent on the
cosmological model, due to differences in the angular diameter versus
redshift relation; however, the changes from the standard $\om = 1$,
$\oml = 0$ model are generally small, $\lta\,0.1$ mag, and so the data
were converted to 63.9\,kpc in the $\om =1$, $\oml = 0$ model, and the
model--to--model variations were included in the predicted model
tracks.

The 63.9\,kpc aperture was chosen since this had been adopted for the 6C
observations wherever possible. In cases where there was a nearby
companion galaxy, or other problems prevented the use of the full
aperture, magnitudes for the 6C galaxies were determined for smaller
apertures and aperture corrections applied by Eales {\it et al.} (1997);  
for galaxies at redshifts $z < 0.6$ they assumed that the radio galaxies
follow an elliptical galaxy growth curve (Sandage 1972), and at redshifts
$z > 0.6$ they assumed that the growth curve for the luminosity $L$ of the
galaxy within radius $r$ could be represented by $L(<r)
\propto\,r^{0.35}$. The K--magnitudes of the 3CR galaxies were originally
determined within a 9 arcsec diameter aperture. At large redshifts, this
is almost equivalent to 63.9 kpc, and so the aperture corrections were
small. Larger aperture corrections were necessary at low redshifts since
the 9 arcsec aperture included only the central regions of the galaxy
(Lilly \& Longair 1984).  A de Vaucouleurs profile for giant ellipticals
was used to correct these magnitudes to the standard diameter, since the
analysis of Best {\it et al.} (1998) showed that this is a good
representation of the variation of the K surface brightness with radius.  
It is also consistent with the radial optical light profile found for low
redshift radio galaxies by Lilly, McLean and Longair (1984).

\begin{figure*}
\vspace{3.8 in}
\begin{center}
\includegraphics{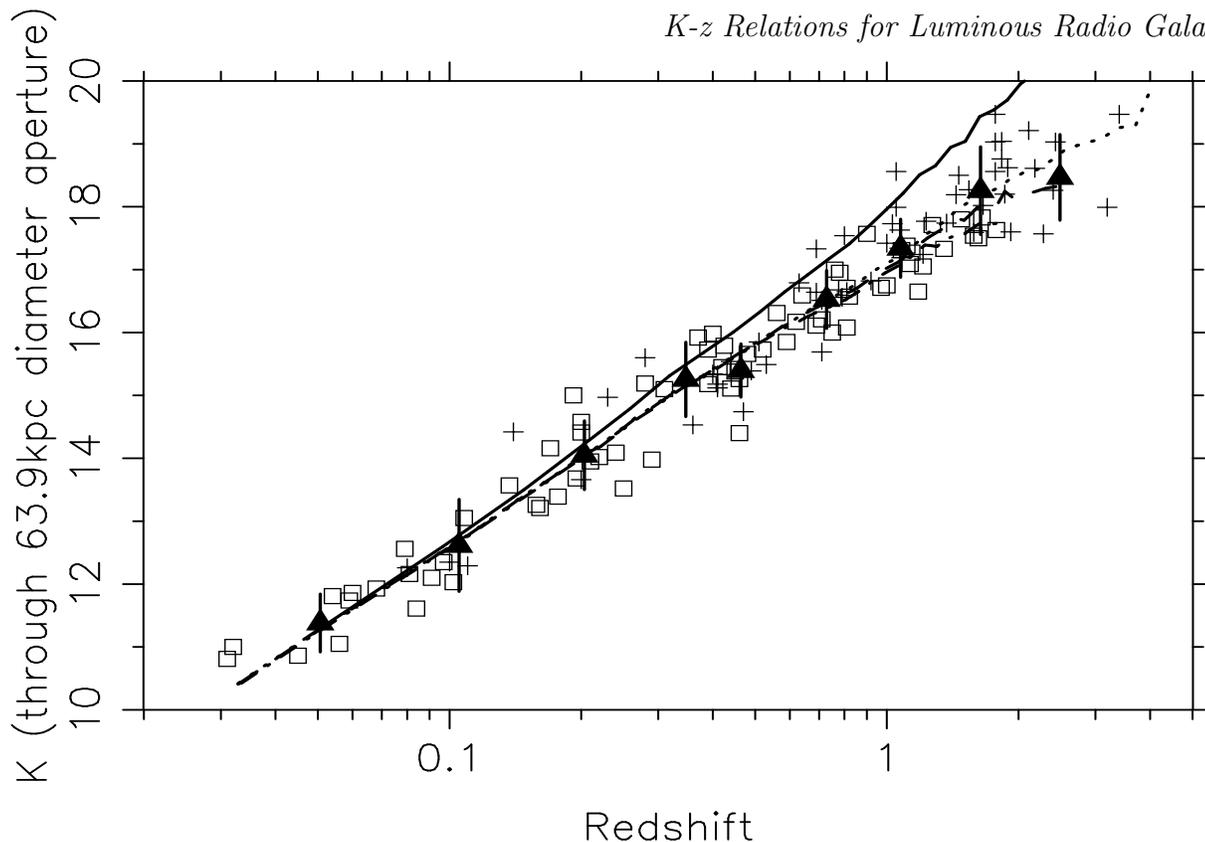}
\end{center}
\caption{The revised K-z diagram for powerful radio galaxies for the
world model with $\om = 1, \oml = 0$.  The open squares represent 3CR
galaxies and the crosses 6C galaxies.  The solid curve corresponds to
the predicted K-z relation for a non-evolving stellar population. The
three other relations are for passively evolving galaxies in which the
stellar population formed at a constant rate for an initial period of
1 Gyr. The three formation redshifts were: $z_{\rm f}= 3$ --
dot-dashed line; $z_{\rm f} = 5$ -- dashed line, and $z_{\rm f} = 20$
-- dotted line. The solid triangles show mean values of the
K-magnitudes for both the 3CR and 6C data averaged in intervals of
$0.1\,T_0$, where $T_0$ is the age of the world model, $T_0 =2/3H_0$
for this case. The bars show the standard deviation of the
distribution of magnitudes in each of these bins.} 
\label{Fig: 1}
\end{figure*}

The resulting K-z relation for 3CR and 6C galaxies, represented by
open boxes and crosses respectively, is shown in Fig.~\ref{Fig: 1}.
New versions of the predicted K-z relations are shown for an $\om =1$,
$\oml = 0$ world model. These involve the use of the most recent set
of spectral synthesis codes (\textsc{gissel}2000), for which a Scalo
Initial Mass Function was adopted, with upper and lower mass cut-offs
of 100M$_\odot$ and 0.1M$_\odot$ respectively; solar metallicity was
assumed. In addition, improved 
K-band transmission data for the K-band filter of UKIRT were used.
Details of these procedures are discussed in Section 3.2. The
differences as compared with the analysis carried out by Best {\it et
al.} (1998) are very small, and so their conclusions for the $\om = 0$
and $\om = 1$ models are unchanged. In particular, the $\om =1$, $\oml
= 0$ model with passive evolution of the stellar content of the
galaxies provides a satisfactory fit to the observations and will be
used as a fiducial model in the new computations.  
    
\subsection{The 3CR, 6C and MRC Radio Galaxies}

One of the intriguing features of the comparison of the 3CR and 6C
data is that, although they are in excellent agreement at redshifts $z
< 0.6$, there is a progressive shift in their mean magnitudes at greater
redshifts, the mean K-magnitude of the 6C galaxies at $z > 1$ being about 0.6
magnitudes fainter on average than the 3CR galaxies (Eales {\it et
al.} 1997). This can be seen
in Fig.~\ref{Fig: 1}, where the boxes, representing 3CR galaxies, lie
towards the low limit of the distribution of the crosses, 6C galaxies,
at $z > 1$.  

\begin{figure}
\vspace{2.35 in}
\includegraphics{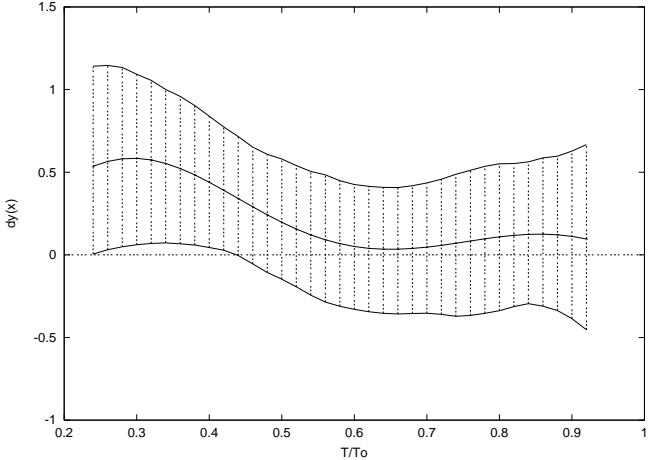} 
\caption{The inferred K-band magnitude difference ({\it dy(x)}) between the 3CR and
6C galaxy populations as a function of cosmic time, expressed as a
fraction of $T_{0}$, the present age of the universe, determined using
a Bayesian statistical approach.  In the Bayesian approach, two flexible distributions for the
reduced magnitude $y(x)$ (the difference between the observed magnitudes
K and the expectation K$_0$ of the no evolution model) are
assumed to underlie the two data sets, for the range of redshifts
covered by both samples.  $y(x)$
is represented by a sum of Legendre polynomials $\Phi$, such that
$y(x)=\Sigma_h w_h \Phi_h(x)$.  Similarly, the scatter in the
observational data $\sigma(x)$ is represented by $\log_{10}\sigma(x) =
\Sigma b_h \Phi_h(x)$. The differences between the two data sets and
their scatter are
defined by the quantities $\delta y(x) = \Sigma \delta w_h \Phi_h(x)$,
and $\delta \log_{10}\sigma(x) = \Sigma
\delta b_h \Phi_h(x)$ respectively.  Gaussian priors were assigned to the
parameters {$w_h$}, {$b_h$}, {$\delta w_h$} and {$\delta b_h$}; the widths of
these four Gaussians were hyperparameters inferred from the
data using the \textsc{bugs} software (Bayesian inference
Using Gibbs Sampling; Thomas {\it et al.} 1992). This method is
similar to that used by MacKay (1995).
The best estimate of the difference
in magnitude between the samples with cosmic time is
plotted as the central solid line. The 95\% confidence limits are
represented by the shaded region, delimited by two solid lines. The
cosmological model used has $\om\,=\,0.3$ and $\oml\,=\,0.7$.} 
\label{Fig: 2}
\end{figure}

Given the quite large dispersion about the mean K-z relation,
Bayesian statistics have been used to investigate the significance of
this magnitude difference and how it evolves with redshift without
making restrictive assumptions about the underlying distribution (see
Figure~\ref{Fig: 2}).   The results show clearly the mean
difference between the two data sets of about 0.6 magnitudes at high
redshift/early times and its decrease to zero at later times. The mean
value of $\delta y(x)$ increases slightly at the lowest redshifts to
about 0.15 magnitudes, but this increase is not significant.  At high
redshifts the difference in K-magnitudes between 3CR and 6C galaxies
is significant at the 95\% confidence limit. 
 
This result for the 6C galaxies is mirrored by that of the still fainter
7C radio galaxies (Willott, priv comm). However, McCarthy (1999) and van
Breugel (1998) have studied the K-z relation for radio galaxies in the MRC
1Jy sample, which contains sources with flux densities $S_{408} > 1$Jy at
408MHz ($\equiv S_{178} \sim 5.8$Jy) from the Molonglo reference catalogue
(McCarthy et al 1996), and they find that the mean K-band magnitudes of
the MRC radio galaxies to be only 0 to 0.2 magnitudes fainter than the
3CR, a much smaller difference than for the 6C sources. The $\sim 25\%$
spectroscopic incompleteness of this sample may introduce a bias in that
it is more difficult to obtain redshifts for intrinsically fainter
galaxies, but McCarthy argues that this bias is unlikely to be
sufficiently great to alter the mean magnitude of the MRC galaxies by 0.6
magitudes.

The important aspect of this discussion is
whether or not there is a correlation between radio luminosity and
K-absolute magnitude, as would be the case if it is accepted that
there is a real difference between the K-z diagrams for the 3CR and 6C
samples.  For the purpose of the present analysis, we have assumed
that the difference between the 6C and 3CR samples, being smaller than
the scatter within either sample, can be ignored. Therefore, we have
taken mean values and dispersions for the combined sample.  These are 
indicated by the solid triangles and associated error bars in
Fig.~\ref{Fig: 1}.    

A literal interpretation of the K-z diagram shown in Fig.~\ref{Fig: 1}
suggests that to a good approximation the luminous radio galaxies
belong to a single population of passively evolving galaxies of the
same stellar mass. 
The galaxies also have the same characteristic radii of about 10\,kpc 
over the entire range $0 < z < 1.5$  (Best et al 1998, McLure \& Dunlop
2000), which supports this conclusion. 
As discussed by Best {\it et al.}, this is a somewhat surprising result, 
since the preferred hierarchical clustering picture of galaxy
formation predicts that the masses and radii of galaxies would grow
significantly over the redshift range considered.   Evidence suggests
that galaxy environments also differ with redshift;
high redshift 3CR galaxies seem to belong to proto-cluster
environments, whereas low redshift galaxies are more commonly
found in small groups or otherwise isolated environments. 
Intuitively, it seems difficult to reconcile these points with
closed--box passive evolution of the stellar populations of the 3CR
galaxies.  
Best {\it et al.} recognised this problem and argued
that the luminous radio galaxies must be observed at a similar stage
in their evolution, specifically that at which they had attained a
certain stellar mass. 
Provided that stars brought together by on--going mergers 
were formed at the same early epoch, the old stellar populations of
galaxies formed by hierarchical clustering would appear to be
passively evolving.
The small scatter in the colour--magnitude relations of elliptical
galaxies observed out to redshifts $z \sim 1$ (Stanford {\it et al.}
1998, Bower {\it et al.} 1998) justifies this assumption.    

A further objective of our analysis is to determine whether it remains
true that the luminous galaxies present on the K-z relation have the
same stellar mass, and how different assumptions about the underlying
cosmology and the evolution of the stellar populations change this
conclusion. 

\section{Modelling Procedures}
\subsection{Cosmological Models}

Because of the availability of the \textsc{gissel} codes of Bruzual
and Charlot, the opportunity has been taken to consider a wider range
of evolutionary histories of the star formation rate of radio
galaxies and to couple this with an expanded range of cosmological
models, in particular, those with non-zero values of the cosmological
constant ($\Lambda$). Friedman world models with zero cosmological
constant and $\Omega_0 = 0, 0.1, 0.3\,{\rm and}\,1.0$ have been
considered, as have a range of models with non-zero $\oml$. The
relation between cosmic time and redshift is: 
\begin{equation}
t=\int_{0}^{t}{\rm d}t = \frac{1}{H_{0}} \int^{\infty}_{z} \frac{{\rm d}z}{(1+z)[\Omega_{0}(1+z)^3 + \Omega_{\Lambda}]^{1/2}}.
\end{equation}
For the world models with flat geometries, $\om + \oml = 1$,
corresponding to those favoured by recent analyses of the power
spectrum of fluctuations in the Cosmic Microwave Background Radiation
as observed in the Boomerang and Maxima experiments (e.g. Jaffe {\it et
al.} 2000), there is a parametric solution:
\begin{equation}
t=\frac{2}{3H_{0}\Omega_{\Lambda}^{1/2}}\ln \left(\frac{1 + \cos\theta}{\sin\theta}\right) 
\qquad
\tan\theta = \left(\dfrac{\Omega_{0}}{\Omega_{\Lambda}}\right)^{1/2}(1+z)^{3/2}. 
\end{equation}
Numerical integrations have been used for other models, and for determining
the variation of co-moving radial distance coordinate with redshift.
The models with non-zero $\oml$ considered in this analysis are listed
in Table 1.  Two of these have flat geometries and the other two
models have open geometries with low values of both $\om$ and $\oml$.
This range of models with finite values of $\oml$ spans those which
would be consistent with the results found in the Supernova Cosmology
project (Perlmutter {\it et al.} 1999).   

\subsection{Galaxy Spectral Synthesis}

The \textsc{gissel} codes of Bruzual and Charlot provide a means of
modelling galaxy spectra through isochrone synthesis.  The models used
in the present analysis all include the passive evolution of the
stellar populations of the galaxies.  The standard model is taken to
have solar metallicity, $Z = Z_{\odot}$, and the stellar population of
the galaxy was formed in an initial uniform starburst with a
duration of 1.0 Gyr. The \textsc{gissel} code enabled the
spectral evolution of this model to be determined for ages up to 20 Gyr
and this is shown in Fig.~\ref{Fig: 3}(a) at a number of ages which
are equally spaced logarithmically in cosmic time. Generally, only the
late stages of evolution of the starburst ($t \gg 10^9$ years) are of
interest in this analysis. For one of the models, $\om=0.1$ and $\oml
= 0.9$, the present age of the Universe, $T_0$, is greater than 20 Gyr
and evolution in this model was terminated at a low redshift
limit of $z = 0.3$.  
 
\begin{table}
\caption{Parameters of the different cosmologies investigated in this
paper. It is assumed that $H_0 = 50$ km s$^{-1}$. Mpc$^{-1}$}  
\begin{center} 
\begin{tabular} {cccc} 
\hline
$\om$ & $\oml$ & \multicolumn{2}{c}{Age at $R=1$} \\
   &   &  (${T_0}\times{H_0}$)& $T_0$ (Gyr)\\  
\hline 
1.0  & 0.0  & 0.67  & 13.0\\
0.3  & 0.0  & 0.81  & 15.8\\
0.1  & 0.0  & 0.90  & 17.6\\
0.0  & 0.0  & 1.00  & 19.6\\\\  
0.3  & 0.7  & 0.96  & 18.9\\
0.3  & 0.3  & 0.86  & 16.9\\
0.1  & 0.9  & 1.28  & 25.0\\
0.1  & 0.3  & 0.98  & 19.1\\
\hline
\end{tabular}
\end{center}
\end{table}

It is instructive to note that, in Fig.~3(a), at optical and infrared
wavelengths the spectra are roughly equally spaced logarithmically in
luminosity at ages greater than $10^9$ years, indicating that, at a 
given wavelength, the spectral energy decreases roughly as a power-law
with cosmic time.  As shown by Gunn (1978), Longair (1998) and others,
this power-law behaviour can be understood in terms of the rate
at which stars evolve from the main sequence onto the red giant branch, if
the initial mass function is assumed to be of power-law form.  Such
simple arguments predict that $L \propto t^{-1}$ for the $\om = 1,
\oml = 0$ model.  This is not so different from the evolution seen in
Fig.~3(a), but there are important variations in the power-law index
as a function of wavelength.  At 0.55 $\mu$m, the evolution can be
roughly described by $L_{0.5} \propto t^{-1.2}$, whereas at 2.2 $\mu$m
the exponent decreases to about $-0.5$. Therefore, the simple
approximation is not adequate for the present analysis. For
essentially all our computations, only evolution at epochs greater
than $10^9$ years is required. It can be seen however that at earlier
times the K-band luminosity increases with age. This is
associated with the evolution of stars onto the asymptotic giant
branch. The stellar evolution models of Maeder and Megnet (1989) and
Girardi {\it et al.} (2000) show that, for stars with $\rm{M} \geq 2
\rm{M}_\odot$, as the stellar mass decreases, the luminosity of the
upper AGB stars increases and their temperature decreases. Thus, the
simple monotonically decreasing spectral luminosity is only applicable
for times greater than $10^9$ years when the turn-off mass drops below
2\,M$_\odot$ and the AGB contribution becomes
less important than evolution onto the red giant branch.

In previous analyses, it was assumed that the metallicity of the
galaxies corresponded to solar values.  The evidence of the Madau plot
in its various incarnations (see, for example, Madau {\it et al.}
1998) and the analyses of the metallicities of the absorption line
systems in distant quasars have suggested that the chemical abundances
of the elements have evolved very significantly over the redshift
interval $0 < z < 4$ (see, for example, Pettini {\it et al.} 1998).
To understand the effects of metallicity upon the predicted K-z
relation, models have been created for four different metallicities:
$0.2 Z_\odot$, $0.4 Z_\odot$, $Z_\odot$, and $2.5 Z_\odot$.  

Different star formation histories have also been considered.  These
include a continuous exponentially decreasing star formation rate,
with an e-folding time of 1 Gyr, and a simple delta-function
starburst.  Combined models could then be constructed involving
multiple star formation events occurring at different cosmic times.
For example, if the onset of radio source and quasar activity were
contemporaneous with (or stimulated) an active starburst, the additional
starburst component 
might affect the K-magnitude of the galaxy. To illustrate how such events
could impact the K-z relation, model starbursts can be added to the
underlying passive evolution of the majority old stellar
population. Two model star bursts were considered. In the first,
the starburst took place instantaneously, and was observed at
times $10^6$, $10^7$ and $10^8$ years after that event; this spectral
evolution is shown in Fig.~\ref{Fig: 3}(b). The intriguing aspect of
these spectra is that the starburst is most luminous in the infrared
waveband after about $10^7$ years. After about this time the most
massive stars complete their evolution on the main sequence and join
the asymptotic giant branch. It is interesting that the time scale of
$10^7$ years is of the same order as the typical age of the radio
source events associated with the 3CR and 6C sources and so such a
star burst could contribute to the K-luminosity of the galaxy. In the
second example, the case of a constant star burst lasting
$10^8$ years is considered, but observed at $10^6$, $10^7$ and $10^8$
years after 
the start of that event; these spectra are shown in Fig.~\ref{Fig:
3}(c). In this case, the star burst is expected to have a much smaller
effect upon the K-luminosity of the galaxy. It can be seen that, after
$10^6$ years, the starburst has the same spectrum as the first model,
but with only one hundredth of the luminosity since only that fraction
of the stars have formed by that time. On the other hand, after $10^8$
years, the starburst is more luminous than in the first model since
there is a significant contribution of recently formed stars to the
K-luminosity. In either case, it can be seen that starbursts observed
between $10^7$ and $10^8$ years after the onset of the starburst can
have a significant impact upon the K-luminosity of the galaxy. Models
have been considered in which the mass fractions of newly formed stars
range between 0.5\% and 5.0\% of the total stellar mass of the galaxy.

\begin{figure}
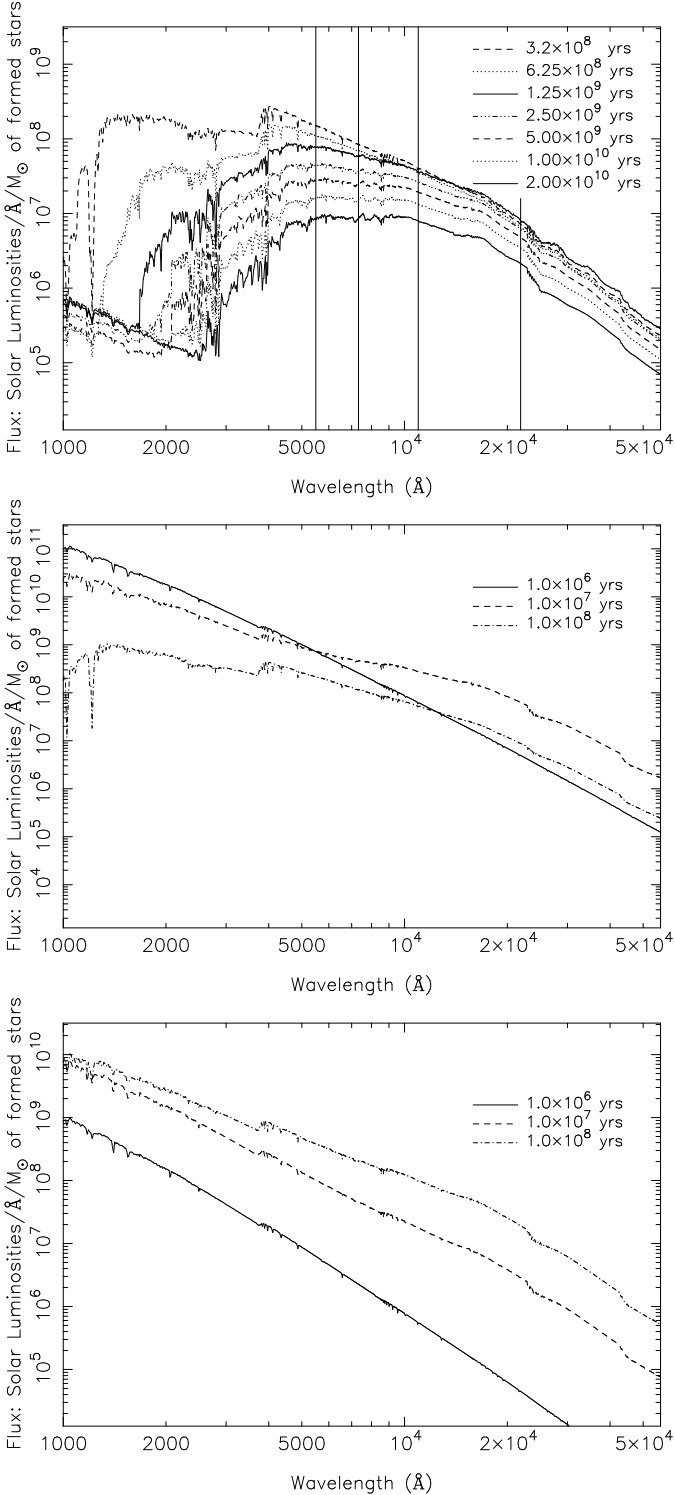

\psfig{file=Fig3a.eps,angle=-90,width=8.9cm,clip=}
\psfig{file=Fig3b.eps,angle=-90,width=8.9cm,clip=} 
\psfig{file=Fig3c.eps,angle=-90,width=8.9cm,clip=}\\
\caption{(a) Synthesised Spectral Energy Distributions for the
standard passive evolution model, consisting of an initial uniform
starburst of duration $10^8$ years with solar metallicity, $Z =
Z_\odot$. The predicted spectral energy distributions of the galaxy is
displayed at eight different ages which are equally spaced
logarithmically in cosmic time. The vertical lines show the regions of
the spectrum corresponding to the rest wavelength of the observed
K-band radiation at redshifts $z\,=\,3,\,2,\,1$ and 0.
(b) SEDs for the additional starbursts
used in our alternative star formation histories, as described in
section 3.2. These are instantaneous starbursts, observed at ages of
$10^6$, $10^7$ and $10^8$ years.  }
\label{Fig: 3}
\end{figure}

\addtocounter{figure}{-1}
\begin{figure}
\caption{\textbf{-- Continued.}   
(c) The spectral evolution of a uniform $10^8$ year starburst observed at ages
$10^6$, $10^7$ and $10^8$ years, the spectrum at the latest time
being the brightest over the full range of wavelength plotted. In all
three figures, the spectra 
are normalised to the luminosity per unit solar mass of stars.}
\end{figure}

The resulting spectral energy distributions (SEDs) were then
redshifted and convolved with the transmission function of the
standard K-filter.  These were then combined with the different world
models to create K-z diagrams. A catalogue of the range of models
considered is given in Tables 3 and 4.  

\section{Results}

\subsection{Changing the Cosmological Model}

\begin{figure}
\vspace{2.6 in}
\includegraphics{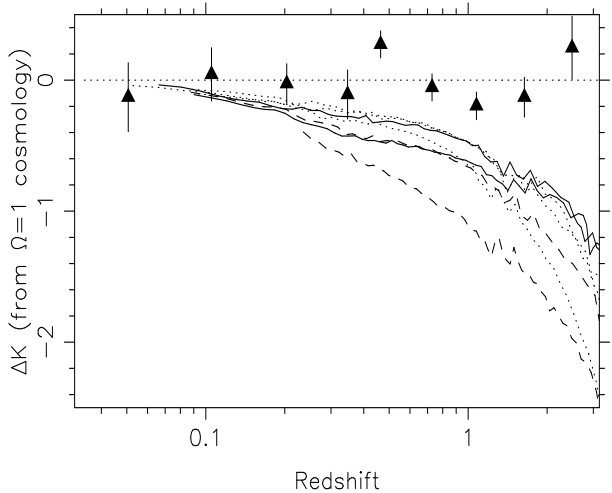}
\caption{Changes in the predicted K-z relations for different
cosmological models. The predictions are presented as differences
relative to that of the $\om = 1$, $\oml = 0$ model. $\Delta {\rm K}$ is
defined as ${\rm K}(\om = 1) - {\rm K(model)}$, so that negative
values for $\Delta {\rm K}$ mean that a galaxy of given absolute
luminosity would be observed to have a fainter apparent magnitude. All
the models 
include evolution of the stellar populations of the galaxies according
to the standard passive evolution model.  Models with $\oml = 0$ and
$\om = 0.0, 0.1, 0.3$ and $1.0$ are shown as dotted lines, the last
model having $\Delta {\rm K} = 0$ by definition; $\Delta {\rm K}$
becomes more negative with decreasing $\om$.  Models with $\om = 0.1$,
$\oml = 0.9$ and $\om =0.1$, $\oml = 0.3$ are shown by dashed lines, the
latter model having the more negative value of $\Delta {\rm K}$.
Models with $\om=0.3$, $\oml = 0.3$ and $\om = 0.3$, $\oml = 0.7$ are
shown by solid lines, the latter model having the more negative value
of $\Delta {\rm K}$.  The triangles and associated error bars
represent the observed values from Fig.~\ref{Fig: 1}.} 
\label{Fig: 4}
\end{figure}

It is simplest to illustrate the effects of changing the parameters of
the cosmological models by subtracting the predicted K-z track for the
chosen cosmology from the corresponding predictions of the $\om=1.0$,
$\oml=0$ model: $\Delta {\rm K}$ is defined as ${\rm K}(\om = 1) -
{\rm K(model)}$ so that negative values of $\Delta {\rm K}$ mean that
a galaxy of a given absolute luminosity would be observed to have a
fainter apparent magnitude.  The results for the cosmologies listed in Table 1
are shown in Fig.~\ref{Fig: 4} and tabulated in Table 2.  

It is instructive to distinguish those contributions to $\Delta {\rm
K}$ due to the
choice of cosmological model and those due to the evolving spectrum of
the galaxy.  The latter can be thought of as the K-correction
including the effects of the evolution of the stellar population.
Following the notation of Longair (1998), the flux density of the
galaxy is given by 
\begin{align}
S_i(\nu_0,z) & = \frac{L[\nu_0(1 + z), t_i(z)]}{4\pi D_i^2(z)(1 + z)}
\end{align} 
where $i$ labels the choice of cosmological model and $D_i(z)$ is a
distance measure which is related to the standard luminosity distance
$D_{\rm L}$ by $D_{\rm L} = D_i(z)(1 + z)$.  
The luminosities of the modelled galaxies in the different cosmologies
are normalised by varying their stellar mass, such that at zero
redshift they have K-band luminosities $L_{\rm{K(z=0)}}$ matching the
observed K-z relation. Therefore if $L^{\prime}[\nu,t]$ is the
luminosity of the galaxy per solar mass,
\begin{equation}
L[\nu_0(1 + z),t_i(z)] = L^{\prime}[\nu_0(1 + z),t_i(z)] \times
\frac{L_{\rm{K(z=0)}}}{L^{\prime}[\nu_0,t_i(0)]} 
\end{equation}
The ratio of the flux densities in the {\it i}th model to
that in the fiducial model, for the same observed flux density at
small redshifts is given by
\begin{equation}
\frac{S_i(\nu_0,z)}{S_0(\nu_0,z)} = \frac{L^{\prime}[\nu_0(1 + z),
t_i(z)]}{L^{\prime}[\nu_0, t_i(0)]}\times \frac{L^{\prime}[\nu_0,
t_0(0)]}{L^{\prime}[\nu_0(1 + z), t_0(z)]} \times \frac{D_0^2(z)}{D_i^2(z)} 
\end{equation}
Thus, the magnitude difference associated with the choice of
cosmological model is
\begin{equation}
\Delta{\rm K}_{\Omega} = 5 \log[D_i(z)/D_0(z)],
\end{equation}
while that associated with observing different parts of the evolving
spectral energy distribution at different cosmic times is  
\begin{equation}
\Delta{\rm K}_{\rm ev} = 2.5 \log\left[\frac{L^{\prime}[\nu_0(1 + z),
t_i(z)]}{L^{\prime}[\nu_0, t_i(0)]}\times \frac{L^{\prime}[\nu_0,
t_0(0)]}{L^{\prime}[\nu_0(1 + z), t_0(z)]}\right]   
\end{equation}
The total change in magnitude is given by $\Delta{\rm K} = \Delta{\rm
K}_{\Omega} + \Delta{\rm K}_{\rm ev}$.  Table~2 shows these
differences for the models listed in Table~1.  

\begin{table*} 
\caption{The cosmic time and distance coordinate $D$, in units of
$c/H_0$ at which galaxies are observed in the cosmological models
listed in Table~2 at redshifts 0, 1, 2 and 3. It is assumed that $H_0
= 50$ km s$^{-1}$ Mpc$^{-1}$.  The third line of each entry shows the
contribution of the choice of cosmological model $\Delta K_{\Omega}$
to $\Delta K$. The fourth line shows the exact contribution of $\Delta
K_{\rm ev}$ to $\Delta K$. The fifth line shows the total value of
$\Delta K$ which is derived from integration over the band
pass of the K-filter.} 
\begin{center} 
\begin{tabular} {cc@{=}lc@{=}r@{.}lc@{=}r@{.}lc@{=}r@{.}lc@{=}r@{.}lc@{=}r@{.}lc@{=}r@{.}lc@{=}r@{.}l} 
\hline
& \multicolumn{2}{c}{$\om = 1$} & \multicolumn{3}{c}{$\om = 0.3$} &
\multicolumn{3}{c}{$\om = 0.1$} & \multicolumn{3}{c}{$\om = 0$} &
\multicolumn{3}{c}{$\om = 0.3$} & \multicolumn{3}{c}{$\om = 0.3$} &
\multicolumn{3}{c}{$\om = 0.1$} &\multicolumn{3}{c}{$\om = 0.1$} \\ 
& \multicolumn{2}{c}{$\oml = 0$} & \multicolumn{3}{c}{$\oml = 0$} &
\multicolumn{3}{c}{$\oml = 0$} & \multicolumn{3}{c}{$\oml = 0$} &
\multicolumn{3}{c}{$\oml = 0.7$} & \multicolumn{3}{c}{$\oml = 0.3$} &
\multicolumn{3}{c}{$\oml = 0.9$} & \multicolumn{3}{c}{$\oml = 0.3$}\\   
\hline 
$z = 0$ & \multicolumn{2}{c}{13.0 Gyr}  & \multicolumn{3}{c}{15.8 Gyr}
& \multicolumn{3}{c}{17.6 Gyr} & \multicolumn{3}{c}{19.6 Gyr}  &
\multicolumn{3}{c}{18.9 Gyr} & \multicolumn{3}{c}{16.9 Gyr} &
\multicolumn{3}{c}{25.0 Gyr} & \multicolumn{3}{c}{19.1 Gyr} \\
& $D$ & 0.00  & $D$& 0 & 00 & $D$& 0 & 00  & $D$& 0 & 00 & $D$& 0 & 00  & $D$& 0 & 00 & $D$& 0 & 00 & $D$& 0 & 00\\
& $\Delta{\rm K}_{\Omega}$ & 0.00& $\Delta{\rm K}_{\Omega}$ & 0 & 00& $\Delta{\rm K}_{\Omega}$ & 0 & 00& $\Delta{\rm K}_{\Omega}$ & 0 & 00& $\Delta{\rm K}_{\Omega}$ & 0 & 00& $\Delta{\rm K}_{\Omega}$ & 0 & 00& $\Delta{\rm K}_{\Omega}$ & 0 & 00 & $\Delta{\rm K}_{\Omega}$ & 0 & 00\\  
&$\Delta{\rm K}_{\rm ev}$ & 0.00&$\Delta{\rm K}_{\rm ev}$ & 0 & 00&$\Delta{\rm K}_{\rm ev}$ & 0 & 00&$\Delta{\rm K}_{\rm ev}$ & 0 & 00&$\Delta{\rm K}_{\rm ev}$ & 0 & 00&$\Delta{\rm K}_{\rm ev}$ & 0 & 00&$\Delta{\rm K}_{\rm ev}$ & 0 & 00&$\Delta{\rm K}_{\rm ev}$ & 0 & 00\\
&$\Delta{\rm K}$ & 0.00 &$\Delta{\rm K}$ & 0 & 00 &$\Delta{\rm K}$ & 0 & 00 &$\Delta{\rm K}$ & 0 & 00 &$\Delta{\rm K}$ & 0 & 00 &$\Delta{\rm K}$ & 0 & 00 &$\Delta{\rm K}$ & 0 & 00 &$\Delta{\rm K}$ & 0 & 00 \\\\
$z = 1$ & \multicolumn{2}{c}{4.61 Gyr}  & \multicolumn{3}{c}{6.54 Gyr} & \multicolumn{3}{c}{7.97 Gyr} & \multicolumn{3}{c}{9.80 Gyr}  & \multicolumn{3}{c}{8.05 Gyr}  & \multicolumn{3}{c}{7.07 Gyr} & \multicolumn{3}{c}{12.73 Gyr} & \multicolumn{3}{c}{8.89 Gyr }\\
& $D$ & 0.59 & $D$ & 0 & 69 & $D$ & 1 & 11 & $D$ & 0 & 75 & $D$ & 0 & 77 & $D$ & 0 & 72 & $D$ & 0 & 90 & $D$ & 0 & 77 \\
& $\Delta{\rm K}_{\Omega}$ & 0.00& $\Delta{\rm K}_{\Omega}$ & $-0$ & 34& $\Delta{\rm K}_{\Omega}$ & $-0$ & 47& $\Delta{\rm K}_{\Omega}$ & $-0$ & 54& $\Delta{\rm K}_{\Omega}$ & $-0$ & 60& $\Delta{\rm K}_{\Omega}$ & $-0$ & 44& $\Delta{\rm K}_{\Omega}$ & $-0$ & 92 & $\Delta{\rm K}_{\Omega}$ & $-0$ & 59\\  
&$\Delta{\rm K}_{\rm ev}$ & 0.00 &$\Delta{\rm K}_{\rm ev}$ & $-0$ & 13 &$\Delta{\rm K}_{\rm ev}$ & 0 & 00 &$\Delta{\rm K}_{\rm ev}$ & $-0$ & 11 &$\Delta{\rm K}_{\rm ev}$ & $-0$ & 02 &$\Delta{\rm K}_{\rm ev}$ & $-0$ & 04 &$\Delta{\rm K}_{\rm ev} $& $-0$ & 21 &$\Delta{\rm K}_{\rm ev}$ & $-0$ & 07\\
&$\Delta{\rm K}$ & 0.00 &$\Delta{\rm K}$ & $-0$ & 47 &$\Delta{\rm K}$ & $-0$ & 47 &$\Delta{\rm K}$ & $-0$ & 65 &$\Delta{\rm K}$ & $-0$ & 62 &$\Delta{\rm K}$ & $-0$ & 48 &$\Delta{\rm K}$ & $-1$ & 13 &$\Delta{\rm K}$ & $-0$ & 66\\\\
$z = 2$ &\multicolumn{2}{c}{2.51 Gyr}  & \multicolumn{3}{c}{4.34 Gyr} & \multicolumn{3}{c}{4.92 Gyr} & \multicolumn{3}{c}{6.53 Gyr}  & \multicolumn{3}{c}{4.52 Gyr} & \multicolumn{3}{c}{4.07 Gyr} & \multicolumn{3}{c}{7.55 Gyr} & \multicolumn{3}{c}{5.45 Gyr}  \\
& $D$& 0.85  & $D$& 1 & 10 & $D$& 1 & 24  & $D$& 1 & 33 & $D$& 1 & 12 & $D$& 1 & 15  & $D$& 1 & 54 & $D$& 1 & 32  \\
& $\Delta{\rm K}_{\Omega}$ & 0.00& $\Delta{\rm K}_{\Omega}$ & $-0$ & 59& $\Delta{\rm K}_{\Omega}$ & $-0$ & 84& $\Delta{\rm K}_{\Omega}$ & $-0$ & 99& $\Delta{\rm K}_{\Omega}$ & $-0$ & 78& $\Delta{\rm K}_{\Omega}$ & $-0$ & 67& $\Delta{\rm K}_{\Omega}$ & $-1$ & 30 & $\Delta{\rm K}_{\Omega}$ & $-0$ & 97\\  
&$\Delta{\rm K}_{\rm ev}$ & 0.00 &$\Delta{\rm K}_{\rm ev}$ & $-0$ & 29 &$\Delta{\rm K}_{\rm ev}$ & $-0$ & 13 &$\Delta{\rm K}_{\rm ev}$ & $-0$ & 39 &$\Delta{\rm K}_{\rm ev}$ & $-0$ & 10 &$\Delta{\rm K}_{\rm ev}$ & $-0$ & 12 &$\Delta{\rm K}_{\rm ev}$ & $-0$ & 40 &$\Delta{\rm K}_{\rm ev}$ & $-0$ & 21\\
&$\Delta{\rm K}$ & 0.00 &$\Delta{\rm K}$ & $-0$ & 88 &$\Delta{\rm K}$ & $-0$ & 97 &$\Delta{\rm K}$ & $-1$ & 38 &$\Delta{\rm K}$ & $-0$ & 88 &$\Delta{\rm K}$ & $-0$ & 79 &$\Delta{\rm K}$ & $-1$ & 70 &$\Delta{\rm K}$ & $-1$ & 18\\\\
$z = 3$ & \multicolumn{2}{c}{1.63 Gyr} & \multicolumn{3}{c}{2.57 Gyr} & \multicolumn{3}{c}{3.46 Gyr} & \multicolumn{3}{c}{4.90 Gyr}  & \multicolumn{3}{c}{2.96 Gyr}  & \multicolumn{3}{c}{2.72 Gyr} & \multicolumn{3}{c}{5.04 Gyr} & \multicolumn{3}{c}{3.80 Gyr} \\
& $D$& 1.00  & $D$& 1 & 43  & $D$& 1 & 68  & $D$& 1 & 88  & $D$& 1 & 49 & $D$& 1 & 46   & $D$& 1 & 98 & $D$& 1 & 77  \\
& $\Delta{\rm K}_{\Omega}$ & 0.00& $\Delta{\rm K}_{\Omega}$ & $-0$ & 77& $\Delta{\rm K}_{\Omega}$ & $-1$ & 13& $\Delta{\rm K}_{\Omega}$ & $-1$ & 37& $\Delta{\rm K}_{\Omega}$ & $-0$ & 86& $\Delta{\rm K}_{\Omega}$ & $-0$ & 82& $\Delta{\rm K}_{\Omega}$ & $-1$ & 48 & $\Delta{\rm K}_{\Omega}$ & $-1$ & 24\\  
&$\Delta{\rm K}_{\rm ev}$ & 0.00 &$\Delta{\rm K}_{\rm ev}$ & $-0$ & 72 &$\Delta{\rm K}_{\rm ev}$ & $-0$ & 42 &$\Delta{\rm K}_{\rm ev}$ & $-0$ & 85 &$\Delta{\rm K}_{\rm ev}$ & $-0$ & 42 &$\Delta{\rm K}_{\rm ev}$ & $-0$ & 37 &$\Delta{\rm K}_{\rm ev}$ & $-0$ & 81 &$\Delta{\rm K}_{\rm ev}$ & $-0$ & 39\\
&$\Delta{\rm K}$ & 0.00 &$\Delta{\rm K}$ & $-1$ & 49 &$\Delta{\rm K}$ & $-1$ & 55 &$\Delta{\rm K}$ & $-2$ & 22 &$\Delta{\rm K}$ & $-1$ & 28 &$\Delta{\rm K}$ & $-1$ & 19 &$\Delta{\rm K}$ & $-2$ & 29 &$\Delta{\rm K}$ & $-1$ & 63 \\
\hline
\end{tabular}
\end{center}
\end{table*}

Fig.~4 shows that galaxies are expected to be fainter at large
redshifts in all the models considered as compared with the model with
$\om = 1$, $\oml = 0$; this means that for the same observed
K-magnitude, the galaxies would have to be intrinsically more luminous.  
There are several effects which contribute to
this decrease. The most important of these is the fact that, for a
given redshift, galaxies are observed at greater radial co-moving
coordinate distances as compared with the model with $\om = 1$, $\oml =
0$.  In addition, the cosmic times at which the spectral energy
distributions are observed are significantly different. What is
important, however, are the relative epochs at which the galaxies are
observed.  It turns out that, in the passively evolving models, the
relative changes in observed luminosity ($\Delta \rm{K}_{\rm ev}$) are
significantly less than the effects of the cosmological model ($\Delta
\rm{K}_{\Omega}$).  This can be understood from 
Table 2 which shows the ages and distance coordinates at which
galaxies at different redshifts are observed in the different
cosmologies listed in Table 1 assuming $H_0 = 50$ km s$^{-1}$
Mpc$^{-1}$; included in this Table are the contributions of
$\Delta \rm{K}_{\Omega}$, the total change in K-magnitude $\Delta \rm{K}$ and
the difference between them $\Delta \rm{K}_{\rm ev}$, which represents
the contribution of the changing spectral energy distribution of the 
galaxies with cosmic epoch.  

Table 2 demonstrates quantitatively the important point that, at all
redshifts, the choice of cosmological model is always much more
important than the evolution corrections within the context of 
passively evolving models for the stellar populations of the galaxies,
so far as the \emph{differences} from the fiducial model are
concerned.   

It is interesting that it is possible to reproduce quite accurately
the results of the exact computations by interpolating the luminosity
changes at the appropriate redshifted wavelength in the template
spectra shown in Fig.~3(a).  It is found that this approach generally
produces results accurate to about 0.1 mag, which is normally accurate
enough to draw useful conclusions.   The value of this approach is
that it enables the investigator to use Fig.~3(a) to work out quite
precisely the effects of the K-corrections for any other world model
with any given value of Hubble's constant and for alternative
assumptions relative to the template spectral evolution shown in Fig.~3(a).   

Poggianti (1997) provides tables of K-corections and
evolutionary corrections for observations in a number of specific
filters, assuming an $\om = 0.45$ cosmology. We have compared these
results with those of our two most similar cosmologies, i.e.  $\om =
1$, $\oml = 0$ and $\om = 0.3$, $\oml = 0$. Despite a different
set of spectral synthesis models, the evolutionary corrections for an
elliptical galaxy formed with an e-folding time of 1 Gyr given by
Poggianti are in good general agreement with our own, suggesting that
the choice of stellar synthesis code does not have a large impact on the
evolutionary corrections determined.  The K--corrections given by
Poggianti for their single cosmology in the K waveband also agree
well with our own results. 

Considering first the range of models shown in Fig.~4, it is evident that
only the passively evolving $\om = 1, \oml = 0$ model provides a reasonable fit
to the data.  The error bars shown in the figure represent the standard
errors in the mean of the K-magnitude of the galaxies, within the
given redshift intervals, and are less than the standard deviations shown in
Fig.~1 by a factor of $\sqrt N$, where $N \sim 10-20$.  

For illustrative purposes, a quantitative comparison between four of the
models plotted in Fig.~\ref{Fig: 4} is given in Table 3. These are the
models with $\om=1.0, \oml=0.0$; $\om=0.1,\oml=0.0$; $\om=0.1, \oml=0.3$;
and $\om=0.3, \oml=0.7$.  In the first part of the table, the magnitude
difference, $\Delta {\rm K}$, between the data points and the no-evolution
tracks at redshifts $z = 0.5, 1.0$ and 2.0 are listed, and in the second
the difference in $\Delta {\rm K}$ between the data points and the 
passive evolution models of Fig.~\ref{Fig: 4} are shown. These data are also
included as the first two entries in Table 4.  The K-z relations have been
normalised to the same K-magnitude at low redshift, for both the passive
evolution and the no-evolution tracks since the mean magnitudes of the
galaxies at $z < 0.3$ are consistent with a very well-defined linear K-z
relation, as can be seen from Fig.~1.   Positive values of $\Delta {\rm K}$
indicate that the predicted magnitudes correspond to brighter galaxies than
are needed to account for the data.  For example, the $\om = 1.0, \oml = 0$
model results in slightly brighter tracks than the mean of the data points
at $z = 2.0$.

The importance of these results is that, in general, except for the
fiducial model, $\om = 1, \oml = 0$, the galaxies are predicted to be
fainter at large redshifts than they are observed to be, even when
passive evolution is taken into account.  This is a somewhat 
unexpected result. First of all, according to the
hierarchical clustering model, galaxies at large redshifts should be less
luminous than those nearby since they contain fewer stars. The present
results go in entirely the wrong direction.  For the most extreme
${T_0}\times{H_0}$ models the galaxies would need to contain about ten times
more stars to account for these differences. Second, it might be
argued that this result would be found if there were a correlation between
radio luminosity and the K-luminosity of the galaxy.  In this case, the
galaxies at large redshift would be expected to be larger than those at
small redshifts.  The analysis of McLure and Dunlop (2000)
showed, however, that the half-width diameters of the luminosity profiles
of the radio galaxies are invariant as a function of redshift, suggesting
that the galaxies have attained a similar evolutionary state so far as
their total stellar populations are concerned.  

Because of these somewhat surprising conclusions, the impact of
changing the assumptions made in the fiducial model in plausible ways
has been analysed in more detail.   

\begin{table*}
\caption{Changes in the K magnitudes, $\Delta {\rm K}$, relative to
the no-evolution model for four representative world models.  The
first part of this table gives the mean differences $\Delta {\rm K}$
between the no-evolution model for these cosmologies and the
observational data for all points within 20\% of three different
redshifts.  The relations have 
been normalised to the mean observed magnitudes at small redshifts.
Positive values indicate that the data points are brighter than the
no-evolution track at that redshift.  The second half of the table
shows the same differences once account is taken of the passive
evolution of the stellar content of the galaxies. For all four cases,
the standard initial starburst is taken have the following parameters:
$z_{\rm f}=10$, metallicity $Z=Z_{\odot}$, $H_0 =
50\,\rm{km\,s}^{-1}{\rm Mpc}^{-1}$, and the duration of the initial
starburst 1 Gyr.  Negative values indicate that the predicted
evolution in K-band magnitude is less than that required by the data,
that is, the modelled track is fainter than the mean of the observed
relation at that redshift.} 
\begin{center}
\begin{tabular}  {ccr@{ $\pm$ }lr@{ $\pm$ }lr@{ $\pm$ }l}
\hline 
Cosmological & Sample & \multicolumn{6}{c}{$\Delta {\rm K}$ from the
no-evolution track at different redshifts}\\
Model & & \multicolumn{2}{c}{z=0.5} & \multicolumn{2}{c}{z=1.0} & \multicolumn{2}{c}{z=2.0} \\
\hline 
$\om\,=\,1.0$, $\oml\,=\,0.0$ & 3CR Data & 0.597 & 0.140 & 1.022 & 0.126 & 1.780 & 0.114\\
 & 6C Data  & 0.706 & 0.097 & 0.433 & 0.145 & 1.136 & 0.233\\  
& Combined Data & 0.646 & 0.089 & 0.742 & 0.115 & 1.222 & 0.210\\
$\om\,=\,0.1, \oml\,=\,0.0$ & Combined Data & 0.778 & 0.091 & 1.110 & 0.118 & 1.945 & 0.220\\
$\om\,=\,0.1, \oml\,=\,0.3$ & Combined Data & 0.712 & 0.091 & 1.076 & 0.119 & 1.925 & 0.220\\
$\om\,=\,0.3, \oml\,=\,0.7$ & Combined Data & 0.789 & 0.091 & 1.101 & 0.117 & 1.758 & 0.214\\
\hline
& Model & \multicolumn{6}{c}{$\Delta {\rm K}$ difference between model and observations}\\
$\om\,=\,1.0, \oml\,=\,0.0$ & standard  initial starburst & -0.177 & 0.089 & 0.162 & 0.115 & 0.336 & 0.210 \\ 
$\om\,=\,0.1, \oml\,=\,0.0$ & standard  initial starburst & -0.303 & 0.091 &-0.262 & 0.118 & -0.696 & 0.220\\
$\om\,=\,0.1, \oml\,=\,0.3$ & standard  initial starburst & -0.352 & 0.091 &-0.310 & 0.119 & -0.799 & 0.220\\ 
$\om\,=\,0.3, \oml\,=\,0.7$ & standard  initial starburst & -0.420 & 0.091 &-0.295 & 0.117 & -0.454 & 0.214\\ 
\hline 
\end{tabular}
\end{center}
\end{table*}

\begin{table*}
\caption{Changes in the K magnitudes, $\Delta {\rm K}$, relative to
the standard passive evolution model in an $\om = 1, \oml = 0$ world
model for the representative world models.  Results are given for
redshifts $z =0.5, 1.0$ and 2.0. Positive values of $\Delta {\rm K}$,
which is defined as as ${\rm K}_0 - {\rm K}_{\rm model}$, indicate
that the new model results in a greater increase in infrared
luminosity as compared with the standard passive evolution model.}
\begin{center}
\begin{tabular}{lcccc}
\hline
Section & & &$\Delta {\rm K}$ from the observational data:&\\
&&z=0.5&z=1.0&z=2.0\\
\hline
&Standard Passive Evolution Model: & -0.18 & 0.16 & 0.34 \\
&$\om=1.0$, $\oml=0.0$, $z_{\rm f}=10$, $Z=Z_\odot$ &&&\\
&1Gyr starburst, $H_0=50\, \rm{km\,s^{-1}Mpc^{-1}}$. &&&\\
\hline
&Change from standard passive evolution model & \multicolumn{3}{c}{$\Delta {\rm K}$ difference from standard passive evolution model:}\\
& & z=0.5 & z=1.0 & z=2.0 \\
\hline
(4.1)&	$\om=0.1$, $\oml=0.0$ & -0.13 & -0.42 & -1.03\\
&	$\om=0.1$, $\oml=0.3$ & -0.18 & -0.47 & -1.14\\
&	$\om=0.3$, $\oml=0.7$ & -0.24 & -0.46 & -0.79\\\\
(4.2)&	$z_{\rm f}$=2   & 0.10   & 0.27 & - \\
&	$z_{\rm f}$=3   & -0.00  & 0.10 & 0.57 \\
&	$z_{\rm f}$=5   & -0.02  & 0.04 & 0.14\\
&	$z_{\rm f}$=10 	& 0.00   & 0.00 & 0.00\\
&	$z_{\rm f}$=20  & -0.04  & -0.05& -0.14\\
&	$z_{\rm f}$=100 & 0.00   & -0.02& -0.12 \\\\
(4.3)&	$H_0\,=\,50\,\rm{km\,s^{-1}Mpc^{-1}}$ & 0.00 & 0.00 & 0.00\\
&	$H_0\,=\,65\,\rm{km\,s^{-1}Mpc^{-1}}$ &-0.02 & 0.04 &-0.01\\
&	$H_0\,=\,100\,\rm{km\,s^{-1}Mpc^{-1}}$&-0.01 &-0.05 & 0.12\\\\
(4.4)&	$Z = 0.2 Z_{\odot}$ & 0.05 & 0.22 & 0.45\\
&	$Z = 0.4 Z_{\odot}$ & 0.09 & 0.10 & 0.32\\
&	$Z = 1.0 Z_{\odot}$ & 0.00 &  0.00 & 0.00\\
&	$Z = 2.5 Z_{\odot}$ & 0.13 & -0.04 & -0.16\\\\
(4.5)&Instantaneous SFR &  -0.04 &  -0.09 & -0.21\\
&1Gyr Constant rate     &  0.00 &  0.00 & 0.00\\
&Exponential SFR        & 0.05 & 0.07 & 0.05\\\\
(4.5)&Extra starburst, 2.5\% mass, 10$^6$years old & 0.04 & 0.07 & 0.17\\
&Extra starburst, 2.5\% mass, 10$^7$years old 	   & 0.32 & 0.28 & 0.22\\
&Extra starburst, 2.5\% mass, 10$^8$years old 	   & 0.04 & 0.04 & 0.05\\\\
(4.5)&Extra starburst, 10$^7$years old, 0.5\% mass & 0.07 & 0.06 & 0.05\\
&Extra starburst, 10$^7$years old, 1.0\% mass      & 0.14 & 0.12 & 0.09\\
&Extra starburst, 10$^7$years old, 2.5\% mass      & 0.32 & 0.28 & 0.22\\
&Extra starburst, 10$^7$years old, 5.0\% mass      & 0.56 & 0.50 & 0.40\\
\hline
\end{tabular}
\end{center}
\end{table*}

\subsection{Star Formation Epoch}

\begin{figure}
\vspace{4.78 in}
\includegraphics{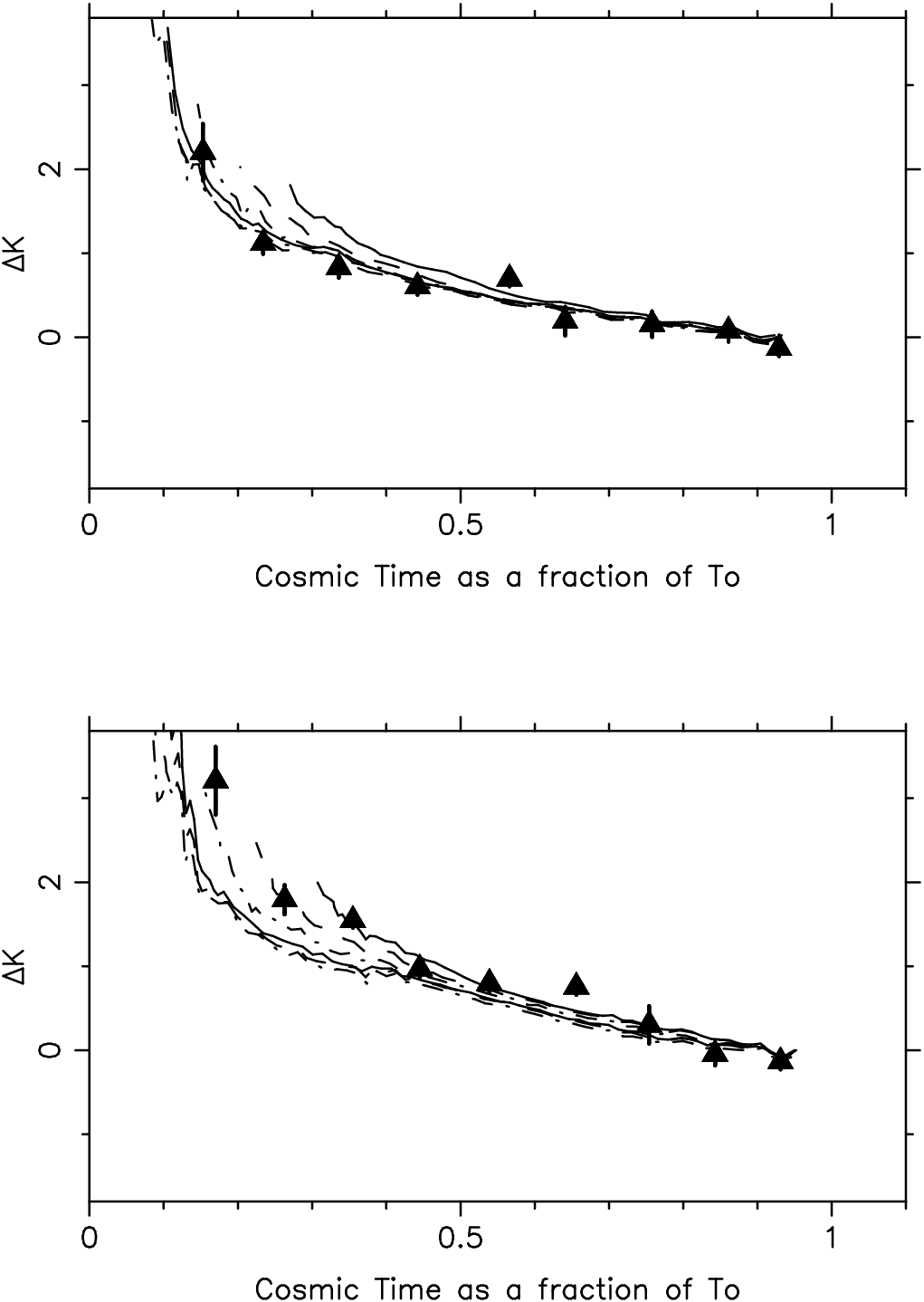} 
\caption{The values of $\Delta {\rm K}$ relative to the no-evolution
model as a function of cosmic time, for galaxies with different
assumed values of the redshifts, $z_{\rm f}$, at which their stellar
populations formed for (a) the critical model, $\om = 1, \oml =0$ and
(b) a flat model with $\om=0.3, \oml=0.7$.  In both plots, upper solid
line corresponds to $z_{\rm f} = 2$, the upper dashed line to $z_{\rm
f} = 3$, the upper dot-dashed line to $z_{\rm f} = 5$, the lower solid
line to $z_{\rm f} = 10$, the lower dashed line to $z_{f}=20$ and
lower dot-dashed line to $z_{\rm f} = 100$.  In both plots it is
assumed that the stars are formed in a 1 Gyr starburst during which
the star formation rate is constant. Solar metallicity is assumed.
The mean values of the observed values of $\Delta {\rm K}$ in equal
cosmic time intervals of $0.1/T_0$ are plotted as in Fig.~\ref{Fig:
1}, with error bars corresponding to the standard error in the
mean of the K-magnitudes of both sets of observational data.}  
\label{Fig: 5}
\end{figure}

The epoch at which the stellar populations were formed was
allowed to vary between redshifts of 2 and 100. The sense of the
changes to the K-z relations are similar for all cosmological
models. Fig.~\ref{Fig: 5} shows the K-band magnitude differences of
the models relative to the no-evolution track in the same cosmology
for models with $\om = 1, \oml = 0$ and $\om = 0.3, \oml =0.7$.  The
magnitude differences relative to the fiducial model are listed in
Table 4 for $\om = 1, \oml = 0$. Results have not been tabulated for
models with $\om = 0.3, \oml =0.7$, as the magnitude differences
within these models were not dissimilar to those in the Einstein--de
Sitter cosmology.  In Fig.~\ref{Fig: 5}, the values of $\Delta {\rm K}$ are
plotted against cosmic time as a fraction of $T_0$, where $T_0$ is the
present age of the Universe listed in Table 1 for $H_0 = 50\,
\rm{km\,s^{-1}Mpc^{-1}}$ .   

In Fig.~\ref{Fig: 5}, it is assumed that the galaxies have solar
metallicity.  All models have been normalised to the same magnitude at
the lowest redshifts, at which the mean K-z relation is well defined.
The models diverge at earlier times. As expected, if the epoch of star
formation is less than the reference value $z_{\rm f} = 10$
the predicted tracks are more luminous at redshifts $z \sim 2 - 3$.
The largest differences occur in the cases in which the 
star formation is assumed to begin at $z_{\rm f} = 2$ and 3, these
being less than that at which radio galaxies are known to
exist.  Nonetheless, it is noteworthy that,
unless the epoch of star formation were as unreasonably low as
$z_{\rm f} = 2 - 3$, changing the epoch of star formation does not result
in a particularly large change in $\Delta {\rm K}$.  The reason for
this can be understood from Fig.~3(a).  By adopting 2.2 $\mu$m as the
wavelength at which the magnitude-redshift relation is determined,
even at redshift $z = 3$, the changes in the luminosity of the
galaxies at a rest wavelength of 500 nm are quite small and change
monotonically with time. Indeed, the difference between observing the
starburst at ages 0.32 to $1.25 \times 10^9$ years amounts to less
than a factor of three at 5000$\rm{\AA}$ in the rest frame. This behaviour
contrasts dramatically with what would be expected if the
redshift-magnitude relation were determined in, say, the R
waveband. Then, Fig.~3(a) shows that the variations in luminosity at
rest frame wavelengths of 1500$\rm{\AA}$ would be very sensitive to even
small amounts of ongoing star formation. These calculations
illustrate the considerable advantages of carrying out studies of the
redshift magnitude relation in the K-waveband -- the observations
provide information more directly about the bulk star
formation activity of the galaxies.  

As expected, as $z_{\rm f}$ increases the predicted K-z relations
become fainter at large redshifts and there is little difference
between the tracks in both models if $z_{\rm f}$ is greater than
10. In the case of the model $\om = 0.3, \oml = 0.7$, the fit to the
data can be improved considerably if the star formation redshift
$z_{\rm f} \sim 3 - 5$ is adopted.  Values of $\Delta {\rm K}$ for the
examples given in Fig.~5(a) at redshifts of 0.5, 1.0 and 2.0 are
included in Table 4.  

\subsection{The Value of Hubble's Constant}

Intuitively, it would seem that the choice of Hubble's constant would
have a significant impact upon the form of the K-z relation since the
cosmic time-scale is significantly different if, say, $H_0$ were are
large as $100\,\rm{km\,s}^{-1}{\rm Mpc}^{-1}$. Then, all cosmic
time-scales would be reduced by a factor of two, allowing much less 
time for the evolution of the stellar populations of the galaxies and
also observing them at very much earlier times at large redshifts.
These changes, as well as those for $H_0 = 65\,\rm{km\,s}^{-1}{\rm
Mpc}^{-1}$, are given in Table 4.  In the cases of models with larger
values for $H_0$, an initial starburst of very short duration was
adopted so that the galaxy is observed during the passive evolution
phase; the tabulated values in Table 4 have included this change in
star formation history. 

It is noteworthy that changing the value of Hubble's constant has a
remarkably small effect upon the predicted evolutionary tracks.  The
reasons for this can be understood from inspection of Fig.~3(a). As was
emphasised in Section 3.2, so long as the analysis is restricted to
redshifts less than 3, the spectral evolution of the stellar energy
distribution is monotonic with time and can be reasonably described by
power-laws at different wavelengths from 0.5 to 2.2 $\mu$m.   Thus, if
the value of Hubble's constant is changed, the stellar mass of
the galaxies required to normalise the K-z relation at low redshift is 
changed, but the differences in the values of $\Delta {\rm K_{ev}}$ depend
primarily upon the changing \emph{relative} cosmic
times observed at different redshifts.  In general, these are quite small
and so changing the value of $H_0$ does not greatly change the evolutionary
corrections.  If the redshift magnitude relation
were determined at rest-frame UV wavelengths, this would certainly not
be the case.

\subsection{Varying the Metallicity of the Galaxies}

\begin{figure}
\vspace{4.78 in}
\includegraphics{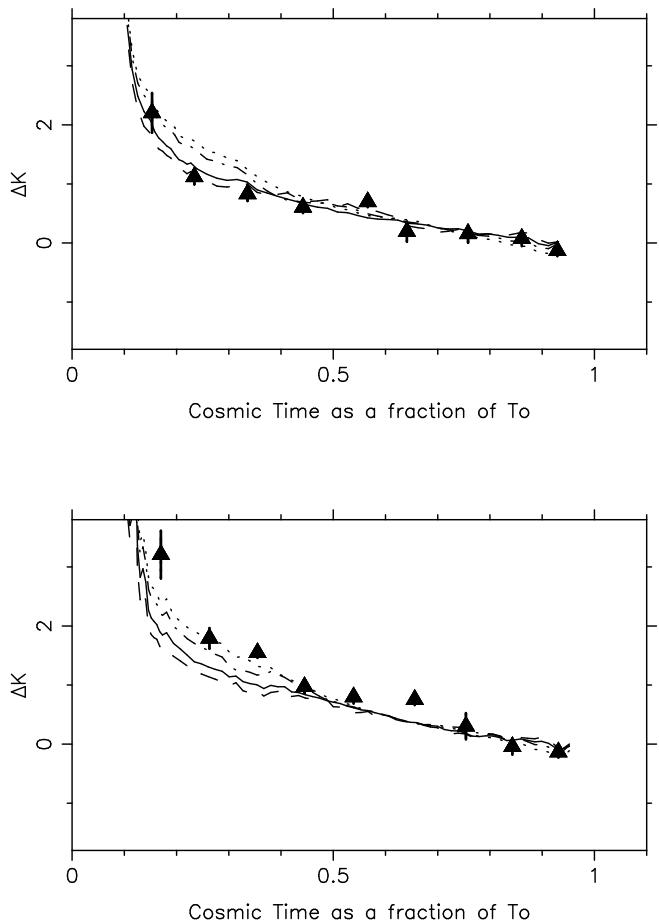}
\caption{The values of $\Delta {\rm K}$ as a function of cosmic time
relative to the no-evolution model for different assumed metallicities
for (a) the critical model, $\om = 1, \oml =0$ and (b) a flat model
with $\om=0.3, \oml=0.7$.  The $\Delta {\rm K}$-cosmic time 
relations for the different metallicities are as follows: dotted line
$Z=0.2Z_{\odot}$; dot-dashed line $Z = 0.4Z_{\odot}$; solid line $Z =
Z_{\odot}$ and dashed line  $Z = 2.5Z_{\odot}$.  The models involve a 1-Gyr
constant starburst, in which star formation begins at $z_{\rm f} = 10$.
Symbols are as in Fig.~\ref{Fig: 5}.}
\label{Fig: 6}
\end{figure}

The metallicity of the galaxies can significantly influence their
evolutionary tracks.  Rather than attempting to adopt some 
arbitrary chemical history, the evolutionary codes have been run for
metallicities of $0.2Z_\odot$, $0.4Z_\odot$, $Z_\odot$ and
$2.5Z_\odot$.  The results are displayed in Fig.~\ref{Fig: 6} for the
same two cosmologies discussed in Section 4.2 and for $z_{\rm f} =
10$.  For the 
Einstein--de Sitter cosmology, these results are also tabulated in
Table 4.  All four K-z tracks are normalised to the same magnitude
scale as the fiducial model, rather than to the same magnitude at the
present epoch. This procedure has been adopted so that at all times
the variations due to changes in metallicity are clearly displayed.
The sense of the changes is that, the lower the metallicity, the more
luminous the galaxy at early epochs.  It is apparent from
Fig.~\ref{Fig: 6}(b), that assuming a lower metallicity at large
redshifts improves the fit of the $\om=0.3, \oml=0.7$ model as
compared with the fiducial model. 

The lowest metallicity track is initially the
brightest (largest $\Delta {\rm K}$), and the $2.5Z_\odot$ track the
faintest, but the tracks cross over at low redshift. The greatest
variation between the tracks occurs at early 
epochs and, as we have seen from Section 3.2, it is at these young
ages that the galaxies undergo the greatest amount of evolution. 
The early evolution of the spectral energy distribution is highly
sensitive to the metallicity of the galaxy because, for a star of a
given mass, changing the metallicity leads to a change in the age
at which it moves from the main sequence onto the AGB; for lower
values of $Z/Z_\odot$ the stars evolve more rapidly since there is less line
blanketing and lower opacity. This has two consequences: firstly, the
more rapid evolution means that lower metallicity stars were more
luminous at early times; secondly, the stars were also shorter
lived and at later times there are fewer stars present in the
older galaxies, which therefore appear fainter than a higher
metallicity galaxy of a similar age.  

It is interesting that the divergence of the different metallicity
tracks occurs when the Universe was roughly half its present age,
corresponding to the redshift at which the K-z relations for the 3CR
and 6C samples begin to diverge. It is just conceivable that such an
effect could account for the difference between these samples, but it
would require those 6C galaxies which are less luminous than the 3CR
galaxies to have higher metallicities.  There is no obvious reason why
greater radio luminosity should be correlated with lower
metallicity of the stellar populations of the radio galaxies.  

The diagrams in Fig.~\ref{Fig: 6} can be used to construct approximate
evolving galaxy models in which the metallicities of the galaxies
increase with cosmic epoch. Such evolution tracks can be found by
changing from one locus to another as the metallicity increases. The
range of models shown in Fig.~\ref{Fig: 6} illustrates how such
changes would alter the predicted K-z relations.

\subsection{Star Formation History}

The results so far have involved an initial starburst during which the
star formation rate has been assumed to be constant. It is important
to consider also models in which there is ongoing star formation, and
in particular those in which the occurrence of a radio source event
might be associated with a major star formation event in the host
galaxy.  This could occur if, for example, the central black hole were
fuelled following some form of galaxy merger.  To understand the
significance of these results, we need to look more closely at the
origin of the K-band flux density. 

The K-band luminosity of evolving galaxies arises from two stellar
populations. Stars with mass M$\gta$2M$_\odot$ evolve beyond core and
shell helium burning to become asymptotic giant stars; those with
M$\lta$2M$_\odot$ become red giant stars forming the old red giant
branch seen, for example, in globular clusters. For the present study,
we are mainly interested in the evolution of the overall stellar
population of the galaxies over cosmic time-scales greater than
$10^9$years. This age corresponds to the main sequence lifetime of a
star of mass 2$\rm{M}_\odot$. Therefore, at ages greater than
$10^9$years, the evolution is dominated by stars evolving onto the
standard red giant branch and the arguments given in section 3.2 can
account for the decreasing luminosity of the galaxy. On the other
hand, if there are epochs of rapid star formation, the bulk of the
near-infrared luminosity is associated with stars on the asymptotic
giant branch. For the most luminous stars, evolution onto the
asymptotic giant branch takes place after a few times
$10^6$years. Figure~\ref{Fig: 3}(b) shows dramatically the large
increase in K-luminosity of an instantaneous starburst when it is
about $10^7$years old associated with the formation of AGB stars. On
the other hand, if the starburst is of longer duration, say
$10^8$years as illustrated in figure~\ref{Fig: 3}(c), the effect is
somewhat diluted by the continued formation of all types of massive
stars. 

\begin{figure}
\vspace{2.04 in}
\includegraphics{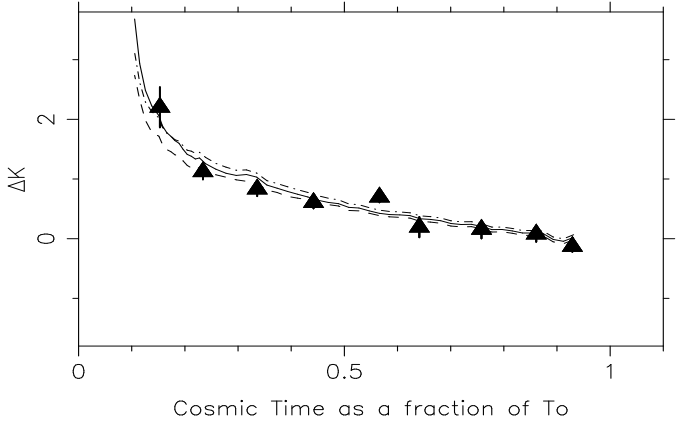}
\caption{The values of $\Delta {\rm K}$ as a function of cosmic time
relative to the no-evolution model for the critical model, $\om = 1,
\oml =0$, for galaxies modelled with different star formation
histories.  Dashed line: instantaneous starburst. Solid line: 1 
Gyr starburst with constant star formation rate. Dot-Dashed line:
Exponentially decaying starburst with $10^9$ year e-folding
time-scale. Galaxies are modelled assuming solar
metallicity, and $z_{\rm f} = 10$. Symbols are as in Fig.~\ref{Fig:
5}(a).}
\label{Fig: 7}
\end{figure}

Considering first variations in the long-term star formation history
of the galaxy populations, Fig.~\ref{Fig: 7} displays three different
types of star formation history: (i) the standard 1-Gyr starburst,
(ii) an instantaneous starburst of negligibly short duration, and
(iii) a model in which the star formation rate decays exponentially
with an e-folding time of $10^9$ years. There is very little
difference between the three models, all three of 
which have been normalised to the same stellar mass at late cosmic
times.   

The brightest model (largest $\Delta {\rm K}$) except at
the very earliest times is the exponential star
formation model, and the faintest the instantaneous burst model. 
These results are not surprising since, it is only
the behaviour after $10^9$years which is relevant to the K-z relation
for $z\,\leq\,3$.  Once star formation ceases, the
continued star formation of the exponential model, and sustained
production of AGB stars (albeit in small numbers), is sufficient to
make this model the brightest track at low redshift. The most evolved
model, the instantaneous starburst, has the lowest
luminosity at low redshift. These differences between the models 
are very small, as can be seen from the values given in Table 4. 

\begin{figure}
\vspace{2.04 in}
\includegraphics{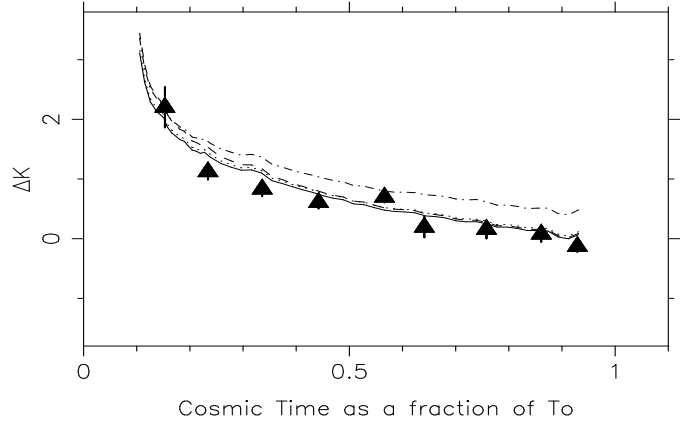}
\caption{Changing $\Delta {\rm K}$ with age of the universe, for
galaxies modelled with complex star formation histories with an
additional starburst occurring a fixed time before the observations,
as outlined in section 2.2. The plotted tracks represent the following
models:  solid line -- no additional starburst, dashed line --
additional starburst occurring $10^6$ years 
prior to the observations, dot-dashed line -- starburst $10^7$ years old,
dotted line -- starburst $10^8$ years old.  Galaxies are modelled with
solar metallicity, and $z_{\rm f} = 10$, in an Einstein-de Sitter
universe.  The extra starburst accounts for 2.5\% of the total stellar
mass of the galaxy.  Symbols are as in Fig.~\ref{Fig: 5}.} 
\label{Fig: 8}
\end{figure}

\begin{figure}
\vspace{2.04 in}
\includegraphics{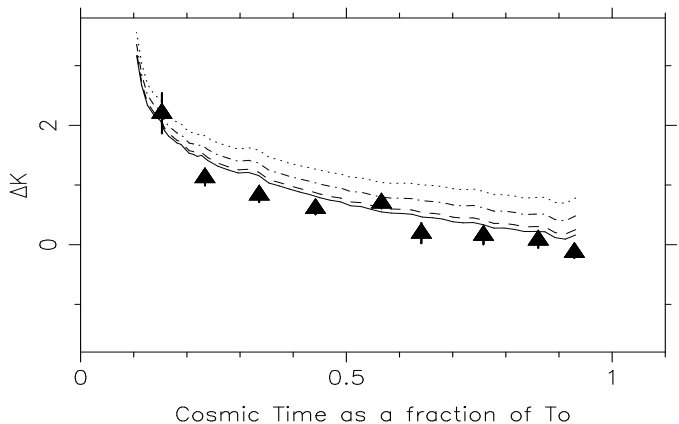}
\caption{Changing $\Delta {\rm K}$ with age of the universe, for
galaxies modelled with star formation histories similar to those in
Fig.~\ref{Fig: 8}.  The plotted tracks represent models with the
following stellar mass fractions involved in the extra starburst:
solid line -- 0.5\%, dashed line -- 1.0\%, dot-dashed line -- 2.5\%,
dotted line -- 5.0\%.  Galaxies are modelled with solar metallicity,
and $z_{\rm f}=10$, in an Einstein--de Sitter universe.  The extra
starburst is modelled as occurring $10^7$ years before the observations
were made.  Symbols are as in Fig.~\ref{Fig: 5}.} 
\label{Fig: 9}
\end{figure}

Fig.~\ref{Fig: 8} shows the standard model coupled with the SEDs of
instantaneous starbursts at three different ages ($10^6$, $10^7$ and
$10^8$ years old), the additional starburst accounting for 2.5\% of
the total stellar mass of the galaxy. The
model with the additional burst occurring the longest time before
observations, $10^8$ years, gives the least change from the
predictions of the underlying model. Similarly, the $10^6$ year old
recent burst also results in only a slight difference at low redshift.  
In the first case, the additional young ($10^8$ years old) high mass
stars have evolved beyond the AGB, and in the second case ($10^6$
years old) they have not yet had enough time for their evolution to reach
that stage. Fig.~\ref{Fig: 3} shows that the rest frame K-band flux of
an evolved instantaneous burst at these ages is much closer to that
for the underlying star formation model than the $10^7$ year old
track. At high redshift, however, we need to consider a somewhat bluer
rest-frame wavelength region. Whilst there is a small difference for
the $10^8$ year old model, the $10^6$ year old recent burst is
slightly brighter still, which is seen in Fig.~\ref{Fig: 8}.  The
model with the $10^7$ year old recent burst component shows the
greatest change from the standard model, as expected from
Fig.~\ref{Fig: 3}. The $\Delta {\rm K}$ values for these models given
in Table 4 are solely due to the additional recent burst
component. The additional burst consistently brightens the K-z tracks
at late times. The reason for the effect being most pronounced at
small redshifts is that the luminosity of the starburst is assumed to
be independent of redshift. Since the K-luminosity increases in
roughly power-law fashion with increasing redshift, the luminosity of
the starburst is soon swamped by the majority population of red giant
branch stars.
Fig.~\ref{Fig: 9} displays the results for the same set of models as
in Fig.~\ref{Fig: 8}, but the mass fraction of the additional
starburst is varied rather than its age. The tracks in this figure
involve an additional $10^7$ year starburst and four mass
fractions.  

If the mass of the starburst as a function of redshift remains
constant, the overall shape of the K-z relation is flattened at late
times and does not result in an improved fit to the observational
data.  Some scatter may be introduced into the K-z relation, because
the radio galaxies are observed at a 
range of ages from $10^6$ to $10^8$ years. In order for starbursts to
change significantly the shapes of the K-z relations, the additional
starburst mass would need to change with redshift. This is quite
possible, as there is more gas available in any merger-induced
starburst at high redshift. The same would be true if the starburst
were radio jet induced (e.g. Rees 1989), which provides a possible
explanation of the alignment effect (McCarthy {\it et al.} 1987,
Chambers {\it et al.} 1987).  Figs.~\ref{Fig: 8} and 9 can be used to
estimate how significantly the K-band luminosity could have changed if
the magnitude of the starburst were to vary with cosmic epoch. Just as
in the case of evolving metallicity, so in the case of evolving
starbursts, the evolutionary history can be estimated by moving from
one track to another in Figs.~\ref{Fig: 8} and 9. For example, in
Fig.~\ref{Fig: 9}, if the early starbursts involved 5\% of the galaxy
mass, the evolutionary track would begin on the dotted line; if this
subsequently decreased to 2.5\% and 1\%, the track would change to the
dot-dashed and dashed lines respectively. A 5\% starburst by mass is
the absolute maximum allowable given the 
constraints of the observed ultraviolet luminosities for the 3CR radio
galaxies at $z \sim 1$ (e.g. Best {\it et al.} 1998). In this extreme case
the magnitude 
increase in K is at most about 0.5 magnitudes.

\section{Discussion and Conclusions}

The above analyses of the K-z relations for a wide range of plausible
cosmological models have established a number of useful general rules
concerning the potential use of the luminosities of massive galaxies
as standard candles.   The value of carrying out these studies in the
K-waveband at 2.2\,$\mu$m for redshifts $z\le 3$ is clearly illustrated
by the synthesised spectral energy distributions shown in
Fig.~\ref{Fig: 3}(a) in which it can be seen that the evolution of the
dominant stellar populations can be predicted with some confidence.
As has been discussed by previous authors, the dominant evolutionary
trends are associated with the evolution of stars in the mass range 1
-- 2 $\rm{M}_{\odot}$ from the main sequence onto the red giant
branch.  An equally important result is that, over this range of
redshifts, the choice of cosmological model is much more important
than the corrections associated the evolution of the stellar
populations of the galaxies, as is vividly illustrated by the values
in Table 2.  Thus, provided standard galaxies can be identified over
the redshift range $0 \le z \le 3$, there are good prospects for
obtaining useful information about the choice of cosmological model.  

The problem with this programme is the identification of such a class
of standard galaxy, particularly in the light of the favoured picture
of hierarchical clustering to account for the formation of structures
such as galaxies and clusters of galaxies over the same redshift
interval.  The analysis of Best {\it et al.} (1998) showed that the
3CR radio galaxies seemed to form a population of galaxies of constant
mass within the framework of the $\om = 1,\oml = 0$ world model,
although they described a number of reasons why this was a somewhat
surprising result.  The present calculations have extended that
analysis to a much wider range of cosmological models and considered
in more detail the effects of the evolution of the stellar populations
of the galaxies. The results shown in Fig.~\ref{Fig: 4} and documented
in Table 2 show that all the new models are a significantly poorer fit
to the large redshift data.  In particular, the principal cause of the
poorer fits is the choice of cosmological model. Expressed in physical
terms, in the critical $\om = 1, \oml = 0$ model, galaxies at a given
redshift are observed at significantly smaller distance measures than
in all the other models considered and so have the brightest
K-magnitudes.   In all the other models considered, a literal
interpretation of the results would be that the radio galaxies at
redshifts $z > 1$ have greater stellar masses than the radio galaxies
nearby.  The analysis has shown that this result is independent of the
choice of Hubble's constant.   

It is of particular interest to find ways in which the favoured flat
cosmological model with $\om =0.3, \oml = 0.7$ can be reconciled with
the observational data. As the results in Table~4 indicate, to produce
an improved fit to the observational data one or more of the following
changes to the standard evolution picture must be introduced:

\begin{enumerate} 

\item[$\bullet$] A lower redshift of star formation, $3 < z_{\rm f} <
4$, provides a much improved fit to the observed K-z relation for this
cosmological model (Fig.~\ref{Fig: 5}b). A problem with this improved
fit is that the galaxies would then be expected to be very much more
luminous in optical wavebands with significant evidence for younger
stellar populations, whilst the red colours of distant radio galaxies
suggest a much larger luminosity--weighted stellar age (e.g. cf Dunlop
{\it et al.}, 1996). Further, the observation of well--formed
galaxies which are powerful radio sources out to redshift $z=5.19$
(van Breugel {\it et al.}, 1999) indicates that at least some of
these galaxies formed at higher redshifts. 

\item[$\bullet$] The metallicities of the galaxies may have been
significantly less at redshifts $z \sim 2$ than they are at the
present epoch.  As shown in Fig.~6(b), decreasing the metallicity of
the galaxy to only about 20\% of the present cosmic value improves
significantly the fit of the model to the observations.

\item[$\bullet$] Starbursts associated with the radio source events
can lead to a significant enhancement of the K-band luminosities of
the galaxies, due to the rapid evolution of luminous massive stars
onto the asymptotic giant branch. This could potentially improve the
fit of the models in lambda cosmologies if associated starbursts
brighten the distant galaxies, but not those at redshifts less than
1. This is quite plausible since such starbursts could be enhanced at
high redshifts due to the extra gas available at these epochs for
forming stars, either in a galaxy merger or triggered by the radio
jet; indeed additional blue luminoisity is observed in powerful radio
galaxies at high redshifts (the alignment effect).  There is clearly
much scope for arbitrarily adjusting the masses, timescales and cosmic
evolution of the starbursts to improve the fit to the K--band
data. However, the excess blue luminosity associated with the
alignment effect requires only a few $\times 10^9 M_{\odot}$ of newly
formed stars ($< 5$\% of the galaxy mass), even were it all to be
associated with star formation activity (e.g. Dunlop {\it et al.}
1989; Best {\it et al.}, 1998): this sets a strong constraint upon
the luminosities and durations of the bursts. 
\end{enumerate}

The alternative approach is to accept that, even after accounting for
stellar evolution, the radio galaxies at $z \ge 1$ may indeed be more
luminous, that is, have a greater stellar mass, than those at small
redshifts. This conclusion would mean that the low and high redshift
radio galaxies would not form a single evolving population, and would
be surprising in terms of hierarchical clustering scenarios according
to which such massive galaxies at high redshift are extremely
rare. Given the growing evidence that the distant radio galaxies
belong to cluster environments, it may be plausible that they had
undergone more mergers than the nearby radio galaxies, which are found
in more rarified environments: this possibility could be tested by
carrying out cross-correlation analyses of radio galaxies of different
K-luminosities with field galaxies, although this becomes a demanding
project for the radio galaxies at $z \ge 1$. The similar
characteristic sizes of the 3CR radio galaxies at redshifts 1 and 0,
however, sets constraints upon the intrinsic differences in
luminosities which effectively rule out the most extreme cosmological
models, those with $T_{\rm 0} \times H_{\rm 0} \gta 1$.

\section*{Acknowledgments}
This work was supported in part by the Formation and Evolution of
Galaxies network set up by the European Commission under contract ERB
FMRX--CT96--086 of its TMR programme. KJI acknowledges the support of
a PPARC research studentship.  PNB would like to thank the Royal
Society for generous financial support through its University Research
Fellowship scheme.

\label{lastpage}

\end{document}